\documentclass[%
	times%
        ,5p%
        ,twocolumn%
	,sort&compress%
	]{elsarticle}

\usepackage{graphicx}
\usepackage{amssymb}
\usepackage[mathscr]{eucal}

\begin{document}

\begin{frontmatter} 
\title{Balancer effects in opinion dynamics}

\author{Taksu Cheon}
\ead[url]{http://researchmap.jp/T\_Zen/} 
\author{Jun Morimoto}

\address{
Laboratory of Physics, Kochi University of Technology, Tosa Yamada, 
Kochi 782-8502, Japan} 

\date{October 30, 2015}

\begin{abstract}
We introduce a novel type of contrarian agent, the balancer, to Galam model of opinion dynamics, in order to account for the existence of social skepticism over one-sidedness.  We find that the inclusion of balancers, along with majoritarian floaters and single-sided inflexibles, brings about the emergence of a critical point on parametric plane of the dynamical system. Around the critical point, three distinct phases of opinion dynamics separated by discontinuous changes are found.
\end{abstract}


\end{frontmatter} 

\section{Introduction}

The opinion dynamics is currently one of the most successful branch of sociophysics \cite{CF09, GA12}.  It is abound with easy-to-analyze dynamical models with intriguing insights for social preferences of people in real-world societies \cite{AA01, AA02, AA03, AA04, AA05, AA06, AA07, AA08, AA09, GA05, GA05a}.   In very simple settings and with simple assumptions, it seems to capture the essence of majority-opinion formation in democratic societies  \cite{BR12, FS14}. The opinion dynamics is formulated in a language of agent-based numerical simulation, but often the equivalent  deterministic dynamical system can be obtained, that can lead to the analytic solutions.

In the Galam model of opinion dynamics, a system with a fixed number of binary-valued agents goes through repeated local-majority updates and reshuffling. 
Two noteworthy findings of the Galam model are the appearance of dominance of minority inflexibles over initial majority \cite{GJ07}, and the persistence of hung election with near fifty-fifty vote \cite{GA04, BG06}.  These findings have been brought about with the introduction of heterogeneous agents to the model:  In addition to  the {\it floater}, the ``normal'' agent type who follows the majority rule, there are two more agent types in the model, {\it inflexible}, and {\it contrarian}.  
Inflexibles are the agents who stick to one opinion whatever the opinions of other agents are.  They can be thought of as representing vested interest, for example.  Inflexibles give rise to the minority-dominance threshold at its population ratio $(3-2\sqrt{2})$ to the dynamics \cite{GJ07}.
Contrarians are the agents who always act contrary to the local majority.  They can be thought of as representing 
skeptical minds concerned with the appearance of unduly powerful majority. Contrarians are found to create the hung election after passing $1/6$ threshold for its ratio among total population \cite{GA04, BG06}.

One curious aspect of original Galam opinion dynamics is that the inflexibles and contrarians mixed together either result in the quick minority dominance of inflexibles, or quick appearance of hung election \cite{JG08}.  This is to be contrasted to the subtler, more varied phenomena  in real-world dynamics of public opinion.
Looking into the political histories of various societies littered with riots and revolutions, we often find that the skeptical few can act to instigate the opposition to oppose the inflexibles, delaying their minority dominance.  There also seems to be occasional ``contrarian overkill'' in which, for example,  independent-minded few help the minority opposition to prevail over the majority supported by the solid inflexibles with vested interest.
These occurrences seem to await a proper modelling in the opinion dynamics.

In this work, we introduce a new agent type, which we call the "balancer", as an alternate modelling of contrarian preferences.   This agent acts as a normal floater except when it meets inflexibles in its updating group, in which occasion, it invariably acts in opposition to the preference of inflexibles.
The conception of the agent with this behavior resulted from the basic observation, that people tend to value fairness in the sense that the society's decision should somehow respect the overall majority.   In any democratic society, we find a rise of people with reasoned skepticism and contrarian attitude, that seek to counter the ``unreasonably'' powerful few.  The contrarians as appeared in the original Galam model, who oppose any majority opinion, are ill-suited to capture such attitude.

The key finding of this work is the uncovering of the critical point in the parameter space formed by population rates of inflexibles and balancers. 
Around the critical point in the parameter space, the system displays very rich dynamics such as the resilience to minority dominance of inflexibles,  the persistent hung election and the balancer overkill.

\section{Opinion dynamics with new element, the balancer}

We consider a system made up of $N$ agents, each of which takes one of two opinions $S$ or $O$ at discrete time $t$, each representing ``support'' or ``oppose'' for a certain issue of common interest to all agents.  We assign a 
binary value $A_t(j)$ to $j$-th agent at time-step $t$, which takes the value 1 for for the opinion $S$ and 0 for $O$.
The opinions of agents are updated deterministically with the discrete time-step advance $t \rightarrow t+1$.   In the update process, the agents are divided into groups of uniform size $r$, and the update is assumed to take place group-locally, that is, the opinion of an agent in time $t+1$ depends only on the opinion of agents sharing the same group at time $t$.  We limit $r$ to an odd integer in this work.  We also assume that $N$ is an integer multiple of $r$, which insures the uniformity of groups.
The central quantity of our interest is the relative size of supporting and opposing agent populations.  We define the 
supporting ratio at time-step $t$ by
\begin{eqnarray}
a_t = \frac{1}{N} \sum_j A_t(j) .
\end{eqnarray}
Obviously, $a_t=1$ signifies the total support where all agents have opinion $S$, and $a_t=0$, the total opposition with all agents having opinion $O$.

The rule of the update is the majority vote with some twists, that come from the heterogeneous characteristics of agents.  We assume that each agent belongs to one of the following {\it agent types}:

\bigskip
1) floater :
This agent updates its opinion always following the majority rule.  It chooses $S$ or $O$ according to the prevailing opinion of the group it belongs. 

\medskip
2) inflexible :  
The opinion of this agent is invariant through all time steps, irrespective to the opinions of others in the group.  There can be both $S$-type inflexibles and $O$-type nflexibles, but in this work, we limit ourselves to the case of the system having only one of the two types.  Since our dynamics is symmetric to $S$ and $O$, the choice is arbitrary, and we only consider inflexibles with invariant opinion $S$, or equivalently, the 
binary value 1.

\medskip
3) balancer :
This agent updates its opinion just like the floaters when there is no inflexibles in the group it belongs, but always updates into the opinion that is opposite to the opinion of inflexibles.  Namely, in the current setting, this agent follows the majority rule in the absence of inflexibles in the group, and takes the opinion $O$ (or equivalently, 0) in its presence.

\bigskip
\noindent{}The last agent type is the new element of the current work.  This type represents a spirit of contrarianism, which acts against opinionated minority wielding excessive influence.  This type of agents are present in all healthy mature democracy.  
%
This type is to be contrasted to the {\it contrarian} type introduced in original Galam model \cite{GJ07} which is characterized by the unconditional opposition to the local majority:

\medskip
3') contrarian :
This agent updates into the state which is counter to the local majority:  It takes the value $O$ at time step $t+1$ if there are more agents with $S$ than ones with $O$ at time step $t$, and takes $S$ at $t+1$ if there are more $O$ than $S$ at $t$. 

\bigskip
\noindent{}We focus on the system consisting of 1) floaters , 2) inflexibles, and 3) balancers in this work, and briefly look at the 
conventional Galam system with 1) floaters , 2) inflexibles, and 3') contrarians for contrasting and  comparison.

After the update, all agents are reshuffled to form new groups for next update.  
We start from a configuration in which the ratio of agents with the opinion $S$ among all agents is $a_0$.  A single update of all agents gives the new supporting ratio $a_1$.  
The procedure is repeated until  the supporting ratio $a_t$ eventually reaches a stable number $a_F$.
This procedure can be viewed either as a model of majority opinion formation in a consensus democracy, or as an idealized description of social decision based on voting in hierarchical representative democracy \cite{GA12}.
\section{Balancer moderation and overkill: Numerical simulations}

Consider a mixed system of floaters, $S$-type inflexibles, and balancers, whose proportion to the total agent population $N$ is $(1-q-b)$, $q$, and $b$, respectively.   We choose the simplest case of smallest nontrivial group size, $r=3$.  
\begin{figure}[ht]
  \centering
  \includegraphics[width=5.5cm]{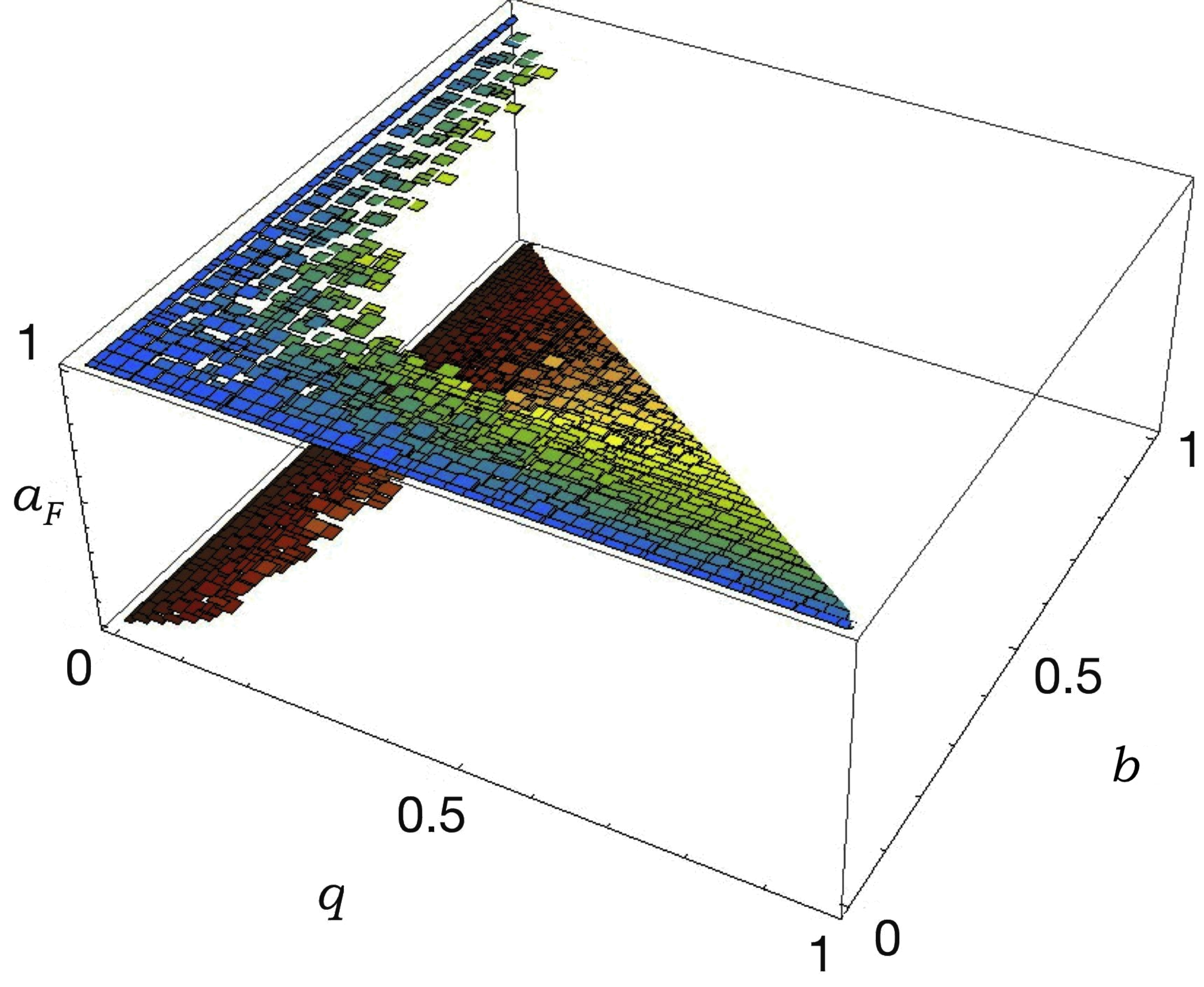}
  \caption{Stable final value of $S$ opinion ratio $a_F$ of the system made up of floaters, inflexibles, and balancers, plotted as a function of inflexible ratio $q$ and balancer ratio $b$: three-dimensional view calculated from numerical simulation. Number of agents are set to $N=240$, and the size of the group is set to $r=3$.  At each run, random initial configuration is evolved for long enough time $T$ to obtain the $s$ ratio $a_T$, which we identify to $a_F$.  The actual value of $T$ is chosen to be $200$, which we have confirmed, by numerically changing $T$, to be large enough.
Note the contrast between the peeled-off structure of the surface in the area $0.3 \gtrsim q \gtrsim 0$ which represents the coexistence of two stable final configurations, and the mono-layered surface in the area $q \gtrsim 0.3$ which represents unique final configuration.  
}
  \label{f1}
\end{figure}
In Figure \ref{f1}, we show the result of numerical simulation with $N=240$, where the stable final configuration $a_F$ is plotted for varying values of $q$ and $b$. 

For small value of $q$ starting from $q=0$, there are two final configurations $a_F \approx q$ and $a_F \approx 1$.
%
There is a sudden disappearance of one of these two final configurations at a certain value of the ratio of inflexibles around $q=0.2\sim 0.3$ as we increase $q$.
Interestingly, the disappearing branch of $a_F$ depends on the value of $b$, the ratio of balancers:  For smaller proportion of balancers $b$, the system with sufficiently large proportion of inflexibles $q$ evolves to a unique final configuration with near total support $a_F \approx 1$. 
The situation is opposite for larger $b$, for which, the system with sufficiently large $q$  evolves to a unique final configuration with $a_F\approx q$ where most of non-inflexible agents have the opinion $O$.
%
%
\begin{figure}[h]
  \centering
  \includegraphics[width=4.0cm]{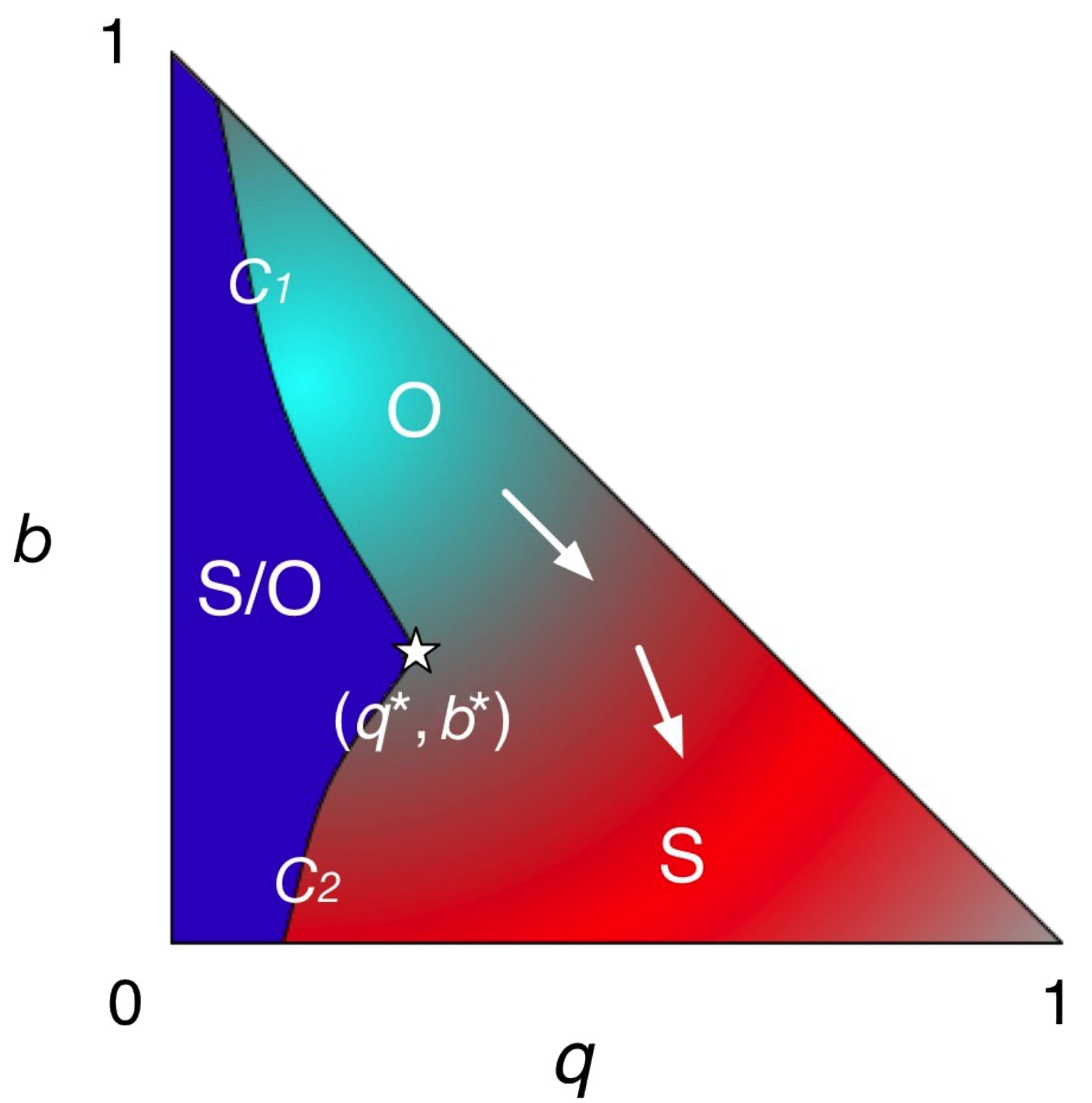}
  \caption{Schematic phase diagram of final configuration $a_F$ of the system made up of floaters, inflexibles, and balancers on the parametric plane of inflexible ratio $q$ and balancer ratio $b$.}
  \label{f2}
\end{figure}

The result could be better visualized with a schematic diagram Figure \ref{f2}.  The whole triangular parameter space ($q\ge 0$, $b\ge 0$, $q+b\le 1$) is split into two regions by two lines $C_1$ and $C_2$, which meet at a branching point $(q^*, b^*)$ located at somewhere around $(\frac{1}{3}, \frac{1}{3})$.  
In $q \approx 0$ side of the lines $C_1$ and $C_2$, two final configurations with $a_F \approx 1$ and $a_F \approx q$ coexist, and a random initial configuration converges toward a final stable configuration with one of these two values, 
signifying the ``democratic'' outcome, in which final majority can go either to $O$ or $S$ depending on the initial configuration.
%
The situation changes when we cross either the line $C_1$ or $C_2$, by increasing the ratio of inflexibles $q$.  After the crossing, only a unique stable configuration exists, signifying the convergence of the distribution of the opinions to a single value of $a_F$ irrespective to the starting population $a_0$.  
%

%
For $b=0$, the $S$-dominance threshold behavior occurs at $q \approx 0.172$, which, of course, is the key finding of the original Galam model \cite{GJ07}.   
As we switch-on the balancer population and increase $b$, the threshold of $S$-dominance caused by inflexible minority is shifted upward.  This can be interpreted as the moderating effect of balancers whose presence works for the restoration of the ``democratic'' majority outcome.  

At the value of $b$ close to $b^*$, the threshold behavior bridging two different phases, as we increase $q$, is softened.  At $b=b^*$, we eventually observe a smooth merging of two final configurations with $a_F \lesssim1/2$ and $a_F \gtrsim1/2$ into a single final configuration of ``hung election''  with $a_F \approx \frac{1}{2}$.

At the value of $b$ above $b^*$, we encounter a novel type of final configuration.   As we increase the inflexible ratio $q$, there is a threshold value of $q$, above which the final configuration with $a_F \approx 1$ disappears, and a unique final configuration of $O$-dominance emerges as a sole outcome.  
This rather surprising result indicates that there can be a {\it dominance of minority against inflexible-supported majority}.  This can be viewed as an ``over-kill effect'' of balancers who work to negate the ``undue influence'' of inflexibles.   This effect might give an explanation for occasional election victory, for example,  by ``progressive'' minority opposition against ``conservative'' majority who are usually backed by vested-interest minority.  
%

\section{Dynamical systems analysis}

Although our system is non-deterministic with random shuffling of group division at each update step, Markovian nature of the dynamics allows us to treat the system, in the limit of infinite agent number $N \to \infty$,  with a deterministic evolution equation for $a_t$, the  ratio of $S$ opinion among all agents at time-step $t$.
%
\begin{table}[ht]
  \centering
  \begin{tabular}{@{} lccccc @{}} 
   \hline 
    $k$ & agents & update   & $\ \ X_k$ & $M_k$ &  $p_k(a)$ \ \ \\
     [0.7ex]
     \hline \\
     [-2.5ex]
    1&  000&{\small 000} $\rightarrow$ {\small 000} & \ \ 0 & 1 & $(1-a-b)^3$ \\
      [1.0ex]
     2& 100&{\small 100} $\rightarrow$ {\small 000} & \ \ 0 & 3 & $(1-a-b)^2 (a-q)$ \\
     3& 110&{\small 110} $\rightarrow$ {\small 111} & \ \ 1 & 3 & $(1-a-b) (a-q)^2$ \\
     4& 111&{\small 111} $\rightarrow$ {\small 111} & \ \ 1 & 1 & $(a-q)^3$ \\
     [0.7ex]
     \cline{2-3} \\
     [-2.5ex]
     5& q00&{\small 100} $\rightarrow$ {\small 100} & \ \ 1/3 & 3 & $(1-a-b)^2 q$ \\
     6& q10&{\small 110} $\rightarrow$ {\small 111} & \ \ 1 & 6 & $(1-a-b) (a-q) q$ \\
     7& q11&{\small 111} $\rightarrow$ {\small 111} & \ \ 1 & 3 & $(a-q)^2 q$  \\
     [1.0ex]
     8& qq0&{\small 110} $\rightarrow$ {\small 111} & \ \ 1 & 3 & $(1-a-b) q^2$  \\
     9& qq1&{\small 111} $\rightarrow$ {\small 111} & \ \ 1 & 3 & $(a-q) q^2$  \\
     10& qqq&{\small 111} $\rightarrow$ {\small 111} & \ \ 1 & 1 & $q^3$  \\
     [0.7ex]
      \cline{2-3} \\
     [-2.5ex]
     11& b00&{\small 000}  $\rightarrow$ {\small 000} & \ \ 0 & 3 & $(1-a-b)^2 b$ \\
     12& b10&{\small 010}  $\rightarrow$ {\small 000} & \ \ 0 & 6 & $(1-a-b) (a-q) b$ \\
     13& b11&{\small 011}  $\rightarrow$ {\small 111} & \ \ 1 & 3 & $(a-q)^2 b$ \\
     [1.0ex]
      \cline{2-3} \\
     [-2.5ex]
     14& bq0&{\small 010} $\rightarrow$ {\small 010}  & \ \ 1/3 & 6 & $(1-a-b) q b$ \\
     15& bq1&{\small 011} $\rightarrow$ {\small 011}  & \ \ 2/3 & 6 & $(a-q) q b$ \\
     16& bqq&{\small 011} $\rightarrow$ {\small 011}  & \ \ 2/3 & 3 & $q^2 b$ \\
     [1.0ex]
     17& bb0&{\small 000} $\rightarrow$ {\small 000} & \ \ 0 & 3 & $(1-a-b) b^2$ \\
     18& bb1&{\small 001} $\rightarrow$ {\small 000} & \ \ 0 & 3 & $(a-q) b^2$ \\
     19& bbq&{\small 001} $\rightarrow$ {\small 001}  & \ \ 1/3 & 3 & $qb^2$ \\
     [1.0ex]
     20& bbb&{\small 000}  $\rightarrow$ {\small 000} & \ \ 0 & 1 & $b^3$ \\
     [0.7ex]
      \hline 
\end{tabular}
\caption{The group agent pattern table for the system with floaters (0/1), inflexibes (q) and balancers (b).
The first column is an indexing label, the second column, all possible formation patterns of agents in a group of size $r=3$ disregarding the order of the appearance, and the third column, the binary value of the agents in that group before and after the update.  The forth column is the probability of obtaining an agent with opinion $S$ after the update,  the fifth column, the multiplicity of the pattern coming from the different orderings, and the sixth column, the probability of the occurrence of the pattern.
}
  \label{tab:booktabs}
\end{table}

For group size three, all possible patterns of agent composition in a group, which we index by an integer $k$, is 20.  Those patterns are tabulated in the second column of Table 1, along with the associated opinion values and their updates, which we list in the third column.  Each of the updated opinion pattern gives contribution $X_k$ to the $S$ ratio after the updates,  which is listed in the 4th column of the Table.   Each pattern $k$ has multiplicity$M_k$, as tabulated in the 5th column, coming from the ordering of heterogenous agents, and each pattern has the probability of appearance $p_k(a)$ for the given $S$ ratio $a$, which is listed in the last column. 

For a system with $q$ inflexibles ratio and $b$ balancer ratio, the update $a_t \longrightarrow a_{t+1}$ is described by
\begin{eqnarray}
a_{t+1} = \sum_{k} X_k \, M_k\,  p_k(a_t) .
\end{eqnarray}
This immediately leads to an explicit form for the evolution equation of $a_t$:
\begin{eqnarray}
a_{t+1} = -2a_t^3 + (3+q) a_t^2 -2q(1+b) a_t +q(1+qb) .\ \ \ 
\label{evol3}
\end{eqnarray}

We have assumed, in the above derivation, that a balancer starts every update with the opinion $O$.  This could be justified for the case of sizeable inflexibles and small number of balancers.
In general, it can also start from $S$, and we will have to expand the Table 1 in order to account for the balancers with both $0$ and $1$ values.  The system, then, is to be described by a set of coupled time-evolution equations with two variables $a_t(f)$ , the $S$ ratio among floaters, and $a_t(b)$, the $S$ ratio among balancers.  These two quantities give the total ratio of $S$ population $a_t=a_t(f)+a_t(b)$ in combination. 
It turns out, however, that our approximate evolution equation (\ref{evol3}) produces surprisingly accurate result compared to the numerical simulation for all values of $q$ and $b$, which should be a sufficient justification of our approximation for now. 

\section{Fixed points of dynamical system}

We can readily obtain the fixed points of the evolution equation (\ref{evol3}) by equating $a_t=a_{t+1} = a_F$.   With straightforward calculation we have%
\begin{eqnarray}
a_F^{[n]} = \frac{3+q}{6}+\frac{\sqrt{\alpha(q,b)}}{3} \cos\left[ \frac{\phi(q,b)-(2n-1)\pi}{3} \right]
\nonumber \\
\qquad(n=1, 2, 3) ,\quad
\label{sol4}
\end{eqnarray}
where we define
\begin{eqnarray}
\phi(q,b) = \arccos \frac{\beta(q,b)}{\sqrt{\alpha(q,b)^3}},
\end{eqnarray}
and
\begin{eqnarray}
&&\alpha(q, b) = q^2-6(1+2b)q+3,
\\ \nonumber
&&\beta(q, b) = -q^3+9(1-4b)q^2-18(1-3b)q.
\end{eqnarray}

To distinguish the stable final configurations and separators of $a_t$ evolution among the solutions $a_F^{[n]}$, it is important to identify regions of parameter space $(q, b)$ in which any of the solutions $a_F^{[n]}$ are real.   On this respect, two sets of dividing lines in the parameter space $(q, b)$ play the central roles.  They are specified by the functions $q=q_C(b)$ and $q=q_A(b)$, which are respectively defined by
\begin{eqnarray}
\alpha(q_C(b), b)^3-\beta(q_C(b), b)^2 = 0,
\label{c1c2}
\end{eqnarray}
and
\begin{eqnarray}
\alpha(q_A(b), b) = 0 .
\end{eqnarray}
Their joint, $\{ q^*, b^*\}$, specified by
\begin{eqnarray}
\alpha(q^*, b^*) = \beta(q^*, b^*) = 0 .
\label{spp}
\end{eqnarray}
is the critical point that characterize the dynamical system.  The actual value of this critical point is obtained as
\begin{eqnarray}
q^*= \frac{3}{8} (7-\eta), \ \   
b^*=\frac{2}{3}\frac{1}{(7-\eta)}-\frac{1}{32}(9+\eta),
\label{sppv2}
\end{eqnarray}
where 
\begin{eqnarray}
\eta = 2\sqrt{17} \cos \left[  \frac{1}{3} \left( \arccos \frac{39}{17\sqrt{17}}-\pi \right) \right] .
\end{eqnarray}
With $\eta \approx 6.198$, we have the numbers, $q^*\approx 0.356327$ and $b^*\approx 0.300747$.
The function $q_C(b)$ can be split into two functions $q_{C_1}(b)$ for $b>b^*$ and $q_{C_2}(b)$ for $b<b^*$.
Suppose we have $b$ below $b^*$. When we increase $q$ starting from $q=0$, there are two real solutions $a_F^{[1]}$ and $a_F^{[3]}$ (the former being larger than the latter), and they merge and disappear when we cross the line $q_{C_2}(b)$.  On the other hand, $a_F^{[2]}$ ($ \leqslant a_F^{[1]}$) persists till it meets the line $q=q_A(b)$, over which it becomes complex, and smoothly taken over by now-real valued $a_F^{[3]}$.

Now suppose we have $b$ above $b^*$.   When we start from $q=0$ and increase $q$, there are two real solutions $a_F^{[2]}$ and $a_F^{[3]}$ (the latter being larger than the former), and they merge and disappear when we cross the line $q_{C_1}(b)$. The other real solution $a_F^{[1]}$ (which is larger than $a_F^{[2]}$) persists until it crosses the line $q=q_A(b)$, over which it becomes complex, and smoothly taken over by $a_F^{[3]}$ which becomes real again beyond that line.  
\begin{figure}[h]
  \centering
  \includegraphics[width=4.2cm]{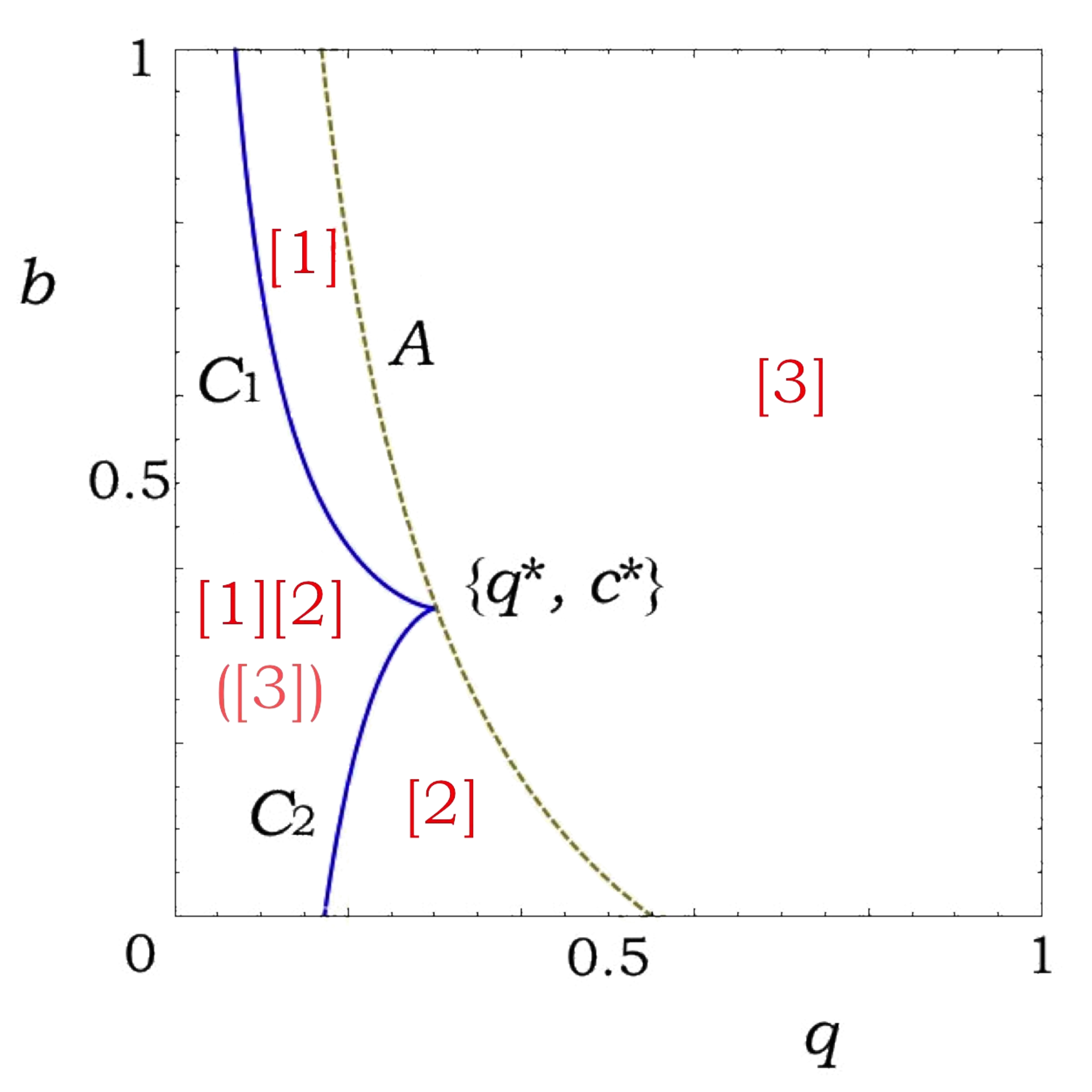}
  \caption{The separating lines specified by $q_A(b)$ and $q_C(b)$, and the critical point $\{ q^*, b^*\}$ of fixed point solution $a_F(q,b)$ depicted on $(q, b)$ plane.  The line $q_C(b)$ has two components $q_{C_1}(b)$ and $q_{C_2}(b)$}
  \label{f3}
\end{figure}
%

The situation is best understood in Figure \ref{f3}, in which we depict the parameter space $(q, b)$ with the the point $\{q^*, d^*\}$ along with the lines defined by $q=q_C(b)$ (which is split into two pieces marked $C_1$ and $C_2$), and by $q=q_A(b)$ (marked $A$).  The index number $[n]$. for which corresponding solution $a_F^{[n]}$ takes a real value in each region delimited by the lines. is also indicated.  The correspondence between the separating lines and the critical point in the numerical analysis, Figure \ref{f2} and the ones in the analytical result, Figure \ref{f3} should be quite evident.

\begin{figure}[h]
  \centering
  \includegraphics[width=4.0cm]{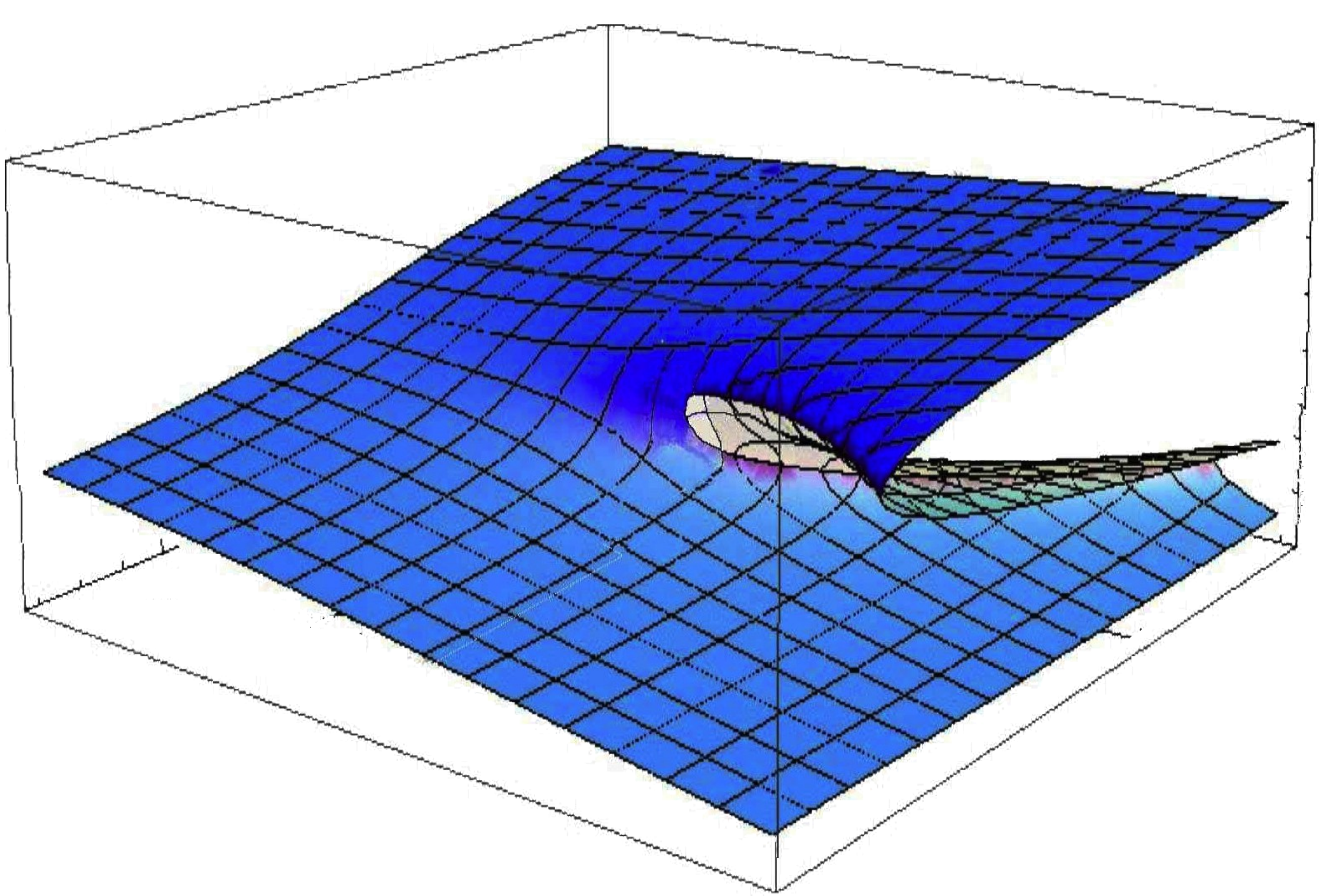}
  \caption{The structure of the fixed point surface $a_F(q,b)$ given by (\ref{sol4}) around the critical point.}
  \label{f4}
\end{figure}
\begin{figure}[h]
  \centering
   \includegraphics[width=4.1cm]{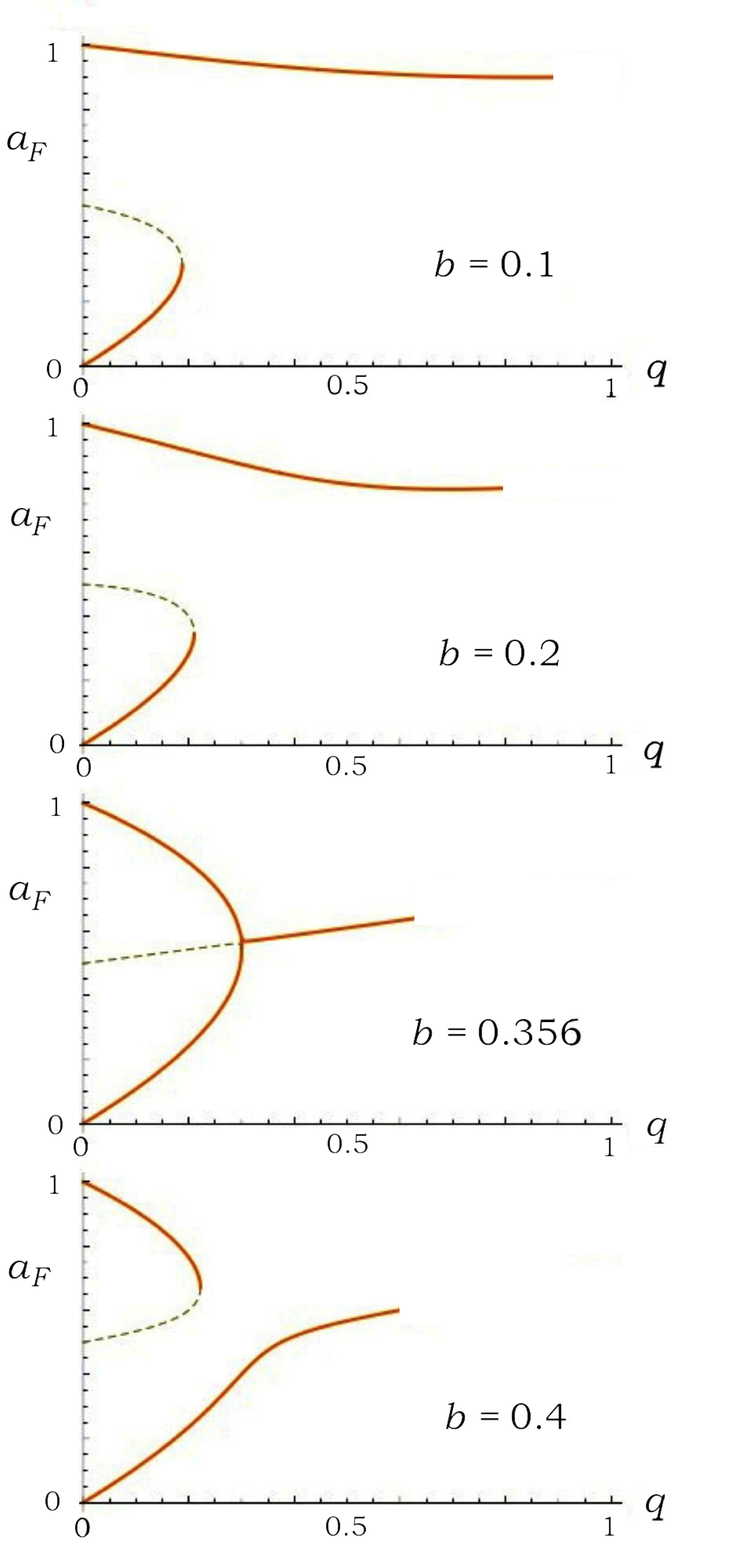}
   \includegraphics[width=4.1cm]{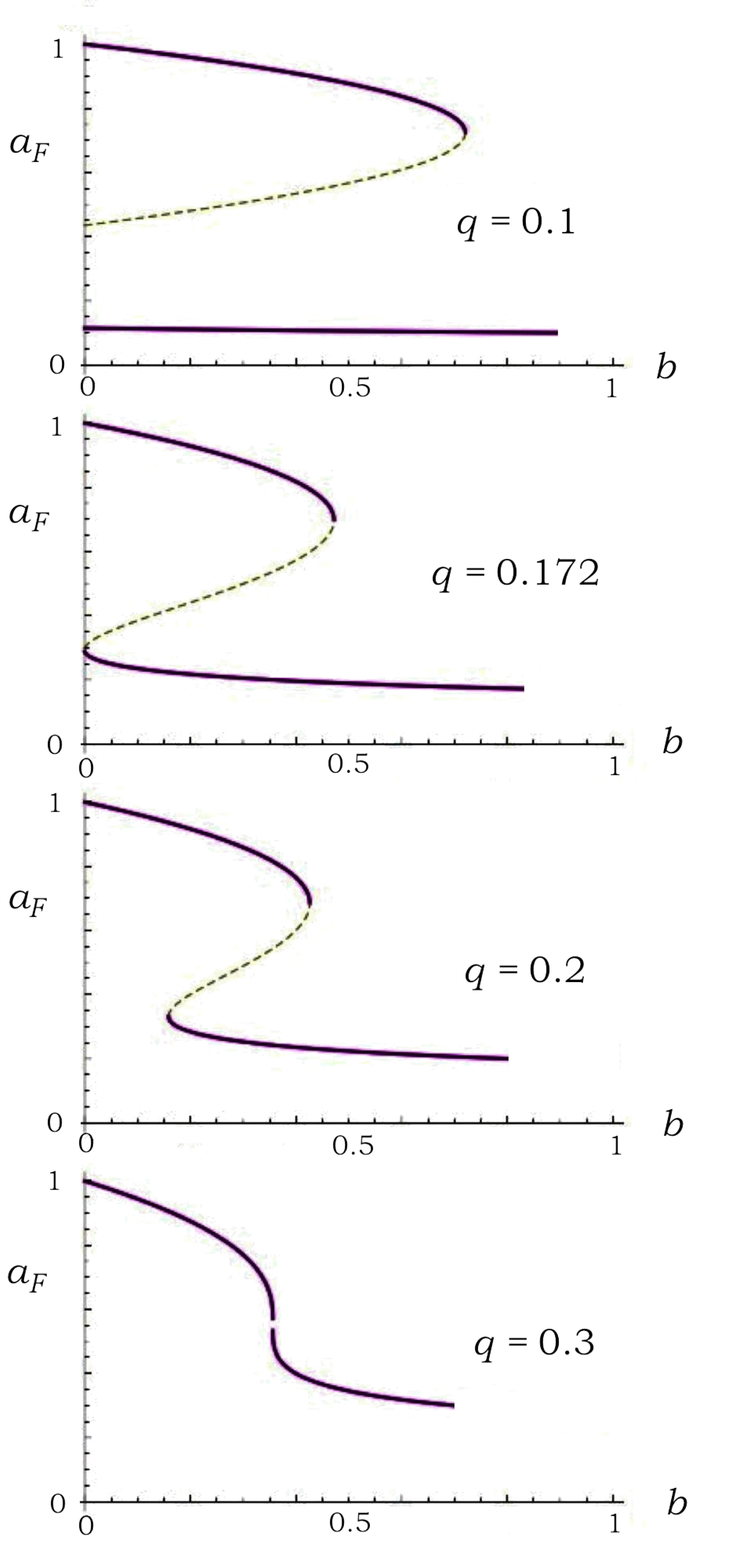}
  \caption{The fixed points and the separators of the system with floaters, inflexbles, and balancers $a_F(q,b)$ given by (\ref{sol4}) shown as functions of inflexible ratio $q$ with a fixed balancer ratio $b$ (left), and as functions of $b$ with a fixed value of $q$ (right).}
  \label{f5}
\end{figure}

The rather complex behavior of the fixed points on the parameter space $(q, b)$ is linked to the intricate branching structure around the critical point $\{q^*, b^*\}$, which can be appreciated by inspecting Figure \ref{f4}, where the surface $a_F(q, b)$ near the critical point is depicted in three-dimensional space $(q, b, a_F)$ . 

\begin{figure}[h]
  \centering
  \includegraphics[width=5.5cm]{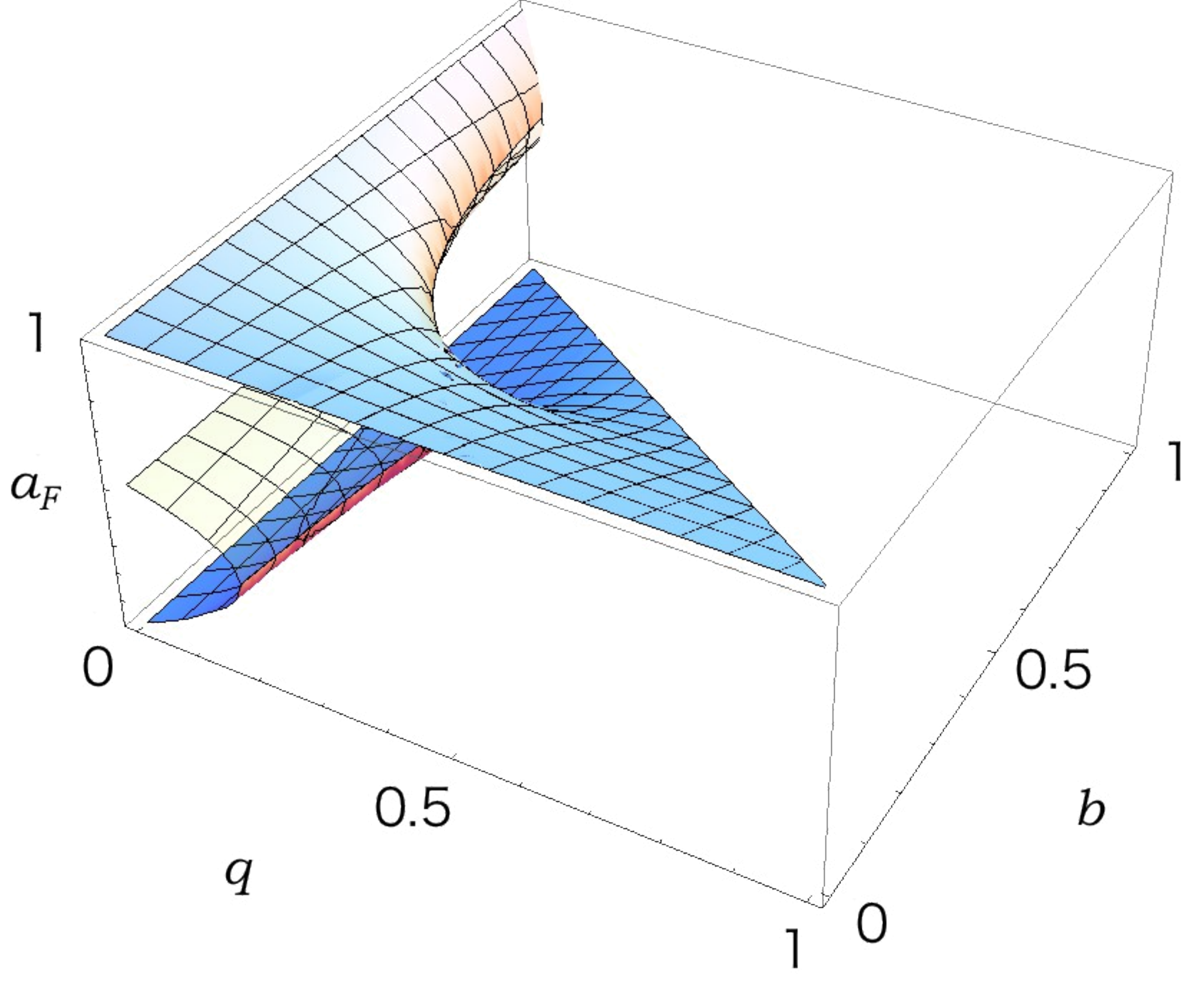}\ 
  \includegraphics[width=5.5cm]{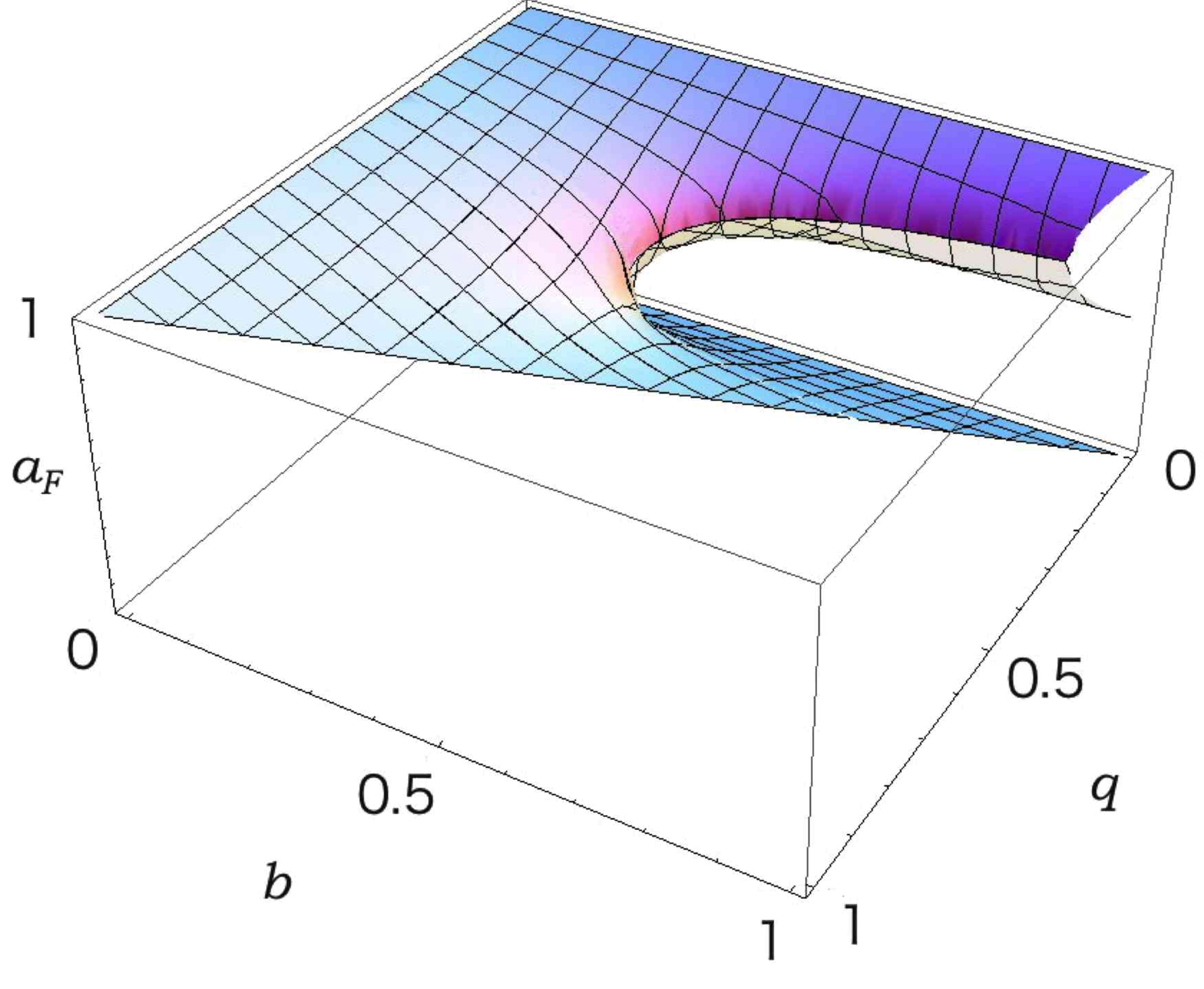}
  \caption{The stable fixed points and separators $a_F(q,b)$ given by (\ref{sol4}) for the system with floaters, inflexbles, and balancers as  functions of inflexible ratio $q$ and balancer ratio $b$ : three-dimensional view.}
  \label{f6}
\end{figure}
To understand the parametric structure of the fixed point more closely, we plot, in Figure \ref{f5},  the stable fixed points and the separator for a given $b$ as functions of $q$ (Figure \ref{f5} left), and  for a given $q$ as functions of $b$ (Figure \ref{f5} right).
Also, in Figure 6, in which we show the three-dimensional plot showing the fixed points as functions on the parameter plane  $\{q, b \}$.
The close resemblance between Figure \ref{f1} and \ref{f6} is a direct evidence of the validity of our approximation neglecting the presence of $O$-values balancer in the process of obtaining (\ref{evol3}).

\begin{figure}[ht]
  \centering
  \includegraphics[width=5.5cm]{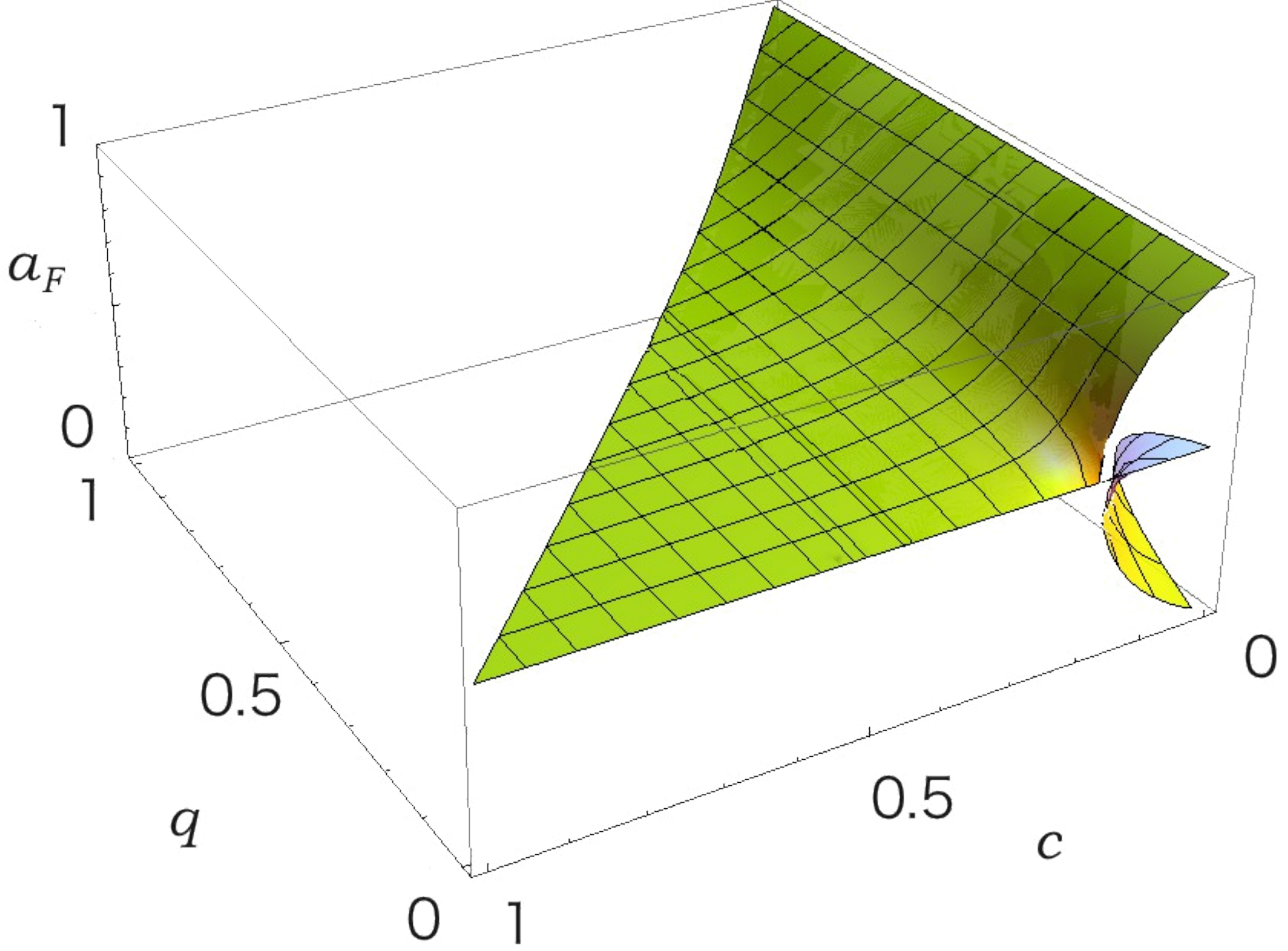}
  \caption{The stable fixed points and separators $a_F(q,b)$ obtained from (\ref{sol12}) for the system with floaters, inflexbles, and contrarians as  functions of inflexible ratio $q$ and contrarian ratio $c$ : three-dimensional view.}
  \label{f7}
\end{figure}
The comparison between our system with balancers and the system with original Galam contrarian is instructive.  From an analogues table to our Table 1 that should be constructed for the contrarian agents, we easily obtain the evolution equation of $a_t$, the $S$ ratio at time-step $t$, for the opinion dynamics with $(1-q-c)$ floater ratio, $q$ inflexible ratio, and $c$ Galam contrarian ratio in the form
\begin{eqnarray}
\label{sol12}
a_{t+1} 
= (-2+4c)a_t^3 + (3+q-6c) a_t^2 -2q  a_t +(q+c) , 
\label{evol12}
\end{eqnarray}
where we have assumed that the contrarian has equal probability of having the opinion $O$ and $S$, an approximation that keeps the symmetry of $O$ and $S$ in the system.

In Figure \ref{f7}, we plot the stable fixed points and the separator on the parameter plane $\{ q, c \}$.  It is obvious that the opinion dynamics of the system with inflexible-contrarian mixture, Figure \ref{f7}, is a ``direct product'' of the model with only inflexibles and the one with only contrarians.  As a result, it is structurally very simple compared to the system with balancers depicted in Figure \ref{f6}.  This difference obviously originates from the distinction between the {\it parametric nonlinearity} of (\ref{evol3}) with respect to $q$ and $b$, and the corresponding linearity of (\ref{evol12}).

\section{Discussions}

If we regard the Galam model as a idealized description of consensus democracy in which people come to social decision after repeated social discourses, it is natural to assume that the agents in the model recognize the types of other agents in the discussion group.   It is harder to imagine such type-recognition of others, if the model is regarded as representing hierarchical majority voting process.  However, voters can often detect the presence of the vested interest and the opinionated extremist minority even in the large-scale voting process, and the assumption of balancer should not be so unnatural.  

Original inspiration of Galam opinion dynamics has come from the spin-chain dynamics in condensed matter physics.
The novel structure found in our model is the result of the interplay between inflexible and balancer agents, that makes the system different from a simple combination of the floater-balancer mixture and the floater-inflexible mixture.
Since the inclusion of a new type of agent, the balancer, has brought interesting effects to the opinion dynamics, it is natural to ask whether similar effects might be found in Ising and other type of spin models in condensed matter physics.  For example, a model with two types of impurities which have certain peculiar mutual interaction might display an analogues dynamics. 
A more natural place to find a counterpart to our balancer dynamics could be the field of epidemics, where immune cells specifically target unfamiliar elements in a system while keeping normal cells unharmed.  
A mathematical treatment of epidemic spreading with immune cells modelled by balancers competing against the cells exposed to disease modelled by inflexible might be of interest to immunologists.

Sociophysical models are still in the early stage of evolution, and models of opinion formation on networks, social, geographical, tele-communicative are now under intensive study.  Inclusion of the balancer type agent in the networked opinion dynamics \cite{AA05, SM04, NM14} is of special interest.
The current model obviously is too simplistic to capture the reality, but this simplicity has allowed us to identify the hidden critical point acting as a lynchpin holding different parametric regions of varying dynamics together.  
With the inclusion of more realistic elements into the model, such as geographical positioning of agents with finite mobility, both temporal and geographical non-uniformity of the system parameters, more than two choices or several coupled two choices, memory effects of agents on past choices,
it may not be too unrealistic to expect that a similar feature is also present in more involved models that aims at precisely describing the dynamics of real-world public opinions.

\bigskip
\noindent{\bf Acknowledgements}

This research was supported by the Japan Ministry of Education, Culture, Sports, Science and Technology under the Grant number 15K05216.



\end{document}